\documentclass[eqsecnum,twocolumn,aps,epsf]{revtex4}

\usepackage{graphicx,amsmath}
\usepackage{color}

\begin{document}


\title{Distinct Fe-induced magnetic states in the underdoped and overdoped regimes of La$_{2-x}$Sr$_x$Cu$_{1-y}$Fe$_y$O$_4$ revealed by muon spin relaxation }

\author{K. M. Suzuki}
\address{Department of Applied Physics, Graduate School of Engineering, Tohoku University, 6-6-05 Aoba, Aramaki, Aoba-ku, Sendai 980-8579, Japan}

\author{T. Adachi}
\thanks{Corresponding author: adachi@teion.apph.tohoku.ac.jp}
\address{Department of Applied Physics, Graduate School of Engineering, Tohoku University, 6-6-05 Aoba, Aramaki, Aoba-ku, Sendai 980-8579, Japan}

\author{Y. Tanabe}
\thanks{{\rm Present address:} WPI-Advanced Institute of Materials Research, Tohoku University, 2-2-1 Katahira, Aoba-ku, Sendai 980-8578, Japan.}
\address{Department of Applied Physics, Graduate School of Engineering, Tohoku University, 6-6-05 Aoba, Aramaki, Aoba-ku, Sendai 980-8579, Japan}

\author{H. Sato}
\address{Department of Applied Physics, Graduate School of Engineering, Tohoku University, 6-6-05 Aoba, Aramaki, Aoba-ku, Sendai 980-8579, Japan}

\author{Risdiana}
\address{Department of Physics, Faculty of Mathematics and Natural Sciences, Padjadjaran University, Jl. Raya Bandung-Sumedang Km. 21 Jatinangor, 45363, Indonesia}
\address{Advanced Meson Science Laboratory, Nishina Center for Accelerator-Based Science, The Institute of Physical and Chemical Research (RIKEN), 2-1 Hirosawa, Wako 351-0198, Japan}

\author{Y. Ishii}
\thanks{{\rm Present address:} Department of Physics, Tokyo Medical University, 6-1-1 Shinju-ku, Tokyo 160-8402, Japan}
\address{Advanced Meson Science Laboratory, Nishina Center for Accelerator-Based Science, The Institute of Physical and Chemical Research (RIKEN), 2-1 Hirosawa, Wako 351-0198, Japan}

\author{T. Suzuki}
\thanks{{\rm Present address:} College of Engineering, Shibaura Institute of Technology, 307 Fukasaku, Minuma-ku, Saitama 337-8570, Japan}
\address{Advanced Meson Science Laboratory, Nishina Center for Accelerator-Based Science, The Institute of Physical and Chemical Research (RIKEN), 2-1 Hirosawa, Wako 351-0198, Japan}

\author{I. Watanabe}
\address{Advanced Meson Science Laboratory, Nishina Center for Accelerator-Based Science, The Institute of Physical and Chemical Research (RIKEN), 2-1 Hirosawa, Wako 351-0198, Japan}

\author{Y. Koike}
\address{Department of Applied Physics, Graduate School of Engineering, Tohoku University, 6-6-05 Aoba, Aramaki, Aoba-ku, Sendai 980-8579, Japan}

\date{\today}

\begin{abstract}

Zero-field and longitudinal-field muon-spin-relaxation measurements have been performed in partially Fe-substituted La$_{2-x}$Sr$_x$Cu$_{1-y}$Fe$_y$O$_4$ in a wide range of hole concentration, to investigate the magnetic state induced by the Fe substitution recently suggested from the neutron-scattering measurements [Phys. Rev. Lett. {\bf 107}, 127002 (2011)].
It has been found that the magnetic transition temperature is notably enhanced through the 1\% Fe substitution in a wide range of hole concentration where superconductivity appears in Fe-free La$_{2-x}$Sr$_x$CuO$_4$.
In the underdoped regime, the Fe-induced magnetic order can be understood in terms of the concept of stripe pinning by Fe as in the case of the Zn-induced one in La$_{2-x}$Sr$_x$Cu$_{1-y}$Zn$_y$O$_4$.
In the overdoped regime, on the other hand, the Fe-induced magnetic order is short-ranged, which is distinct from the stripes.
It is plausible that a spin-glass state of Fe spins derived from the Ruderman-Kittel-Kasuya-Yosida interaction is realized in the overdoped regime, suggesting a change of the ground state from the strongly correlated state to the Fermi-liquid state with hole doping in La-214 high-$T_{\rm c}$ cuprates.

\end{abstract}
\vspace*{2em}
\maketitle
\newpage

\section{Introduction}

Since the discovery of high-$T_{\rm c}$ superconducting (SC) cuprates, the magnetic correlation has resided in the heart of the research of the high-$T_{\rm c}$ superconductivity.
The antiferromagnetic order of Cu spins in parent compounds of the high-$T_{\rm c}$ superconductors (HTSC) disappears with  doping of a small amount of holes, followed by the appearance of the SC state coexisting with the antiferromagnetic fluctuation.
In the underdoped regime of La$_{2-x}$Sr$_x$CuO$_4$ (LSCO), neutron-scattering measurements have revealed that an incommensurate magnetic order relating to the so-called spin-charge stripe order~\cite{Tranquada1995} is formed at low temperatures,~\cite{TSuzuki1998,Fujita2002} while an incommensurate magnetic correlation relating to the dynamical stripe correlations has been observed in a wide doping-range where superconductivity appears in various HTSC.~\cite{A,B,C}
In the overdoped regime of LSCO, even though the magnetic correlation weakens, inelastic neutron-scattering measurements have suggested that the incommensurate magnetic correlation exists and disappears at $x \sim 0.30$ where the superconductivity also disappears.~\cite{Wakimoto2004}
These results impress us with the magnetic correlation deeply related to the appearance of the high-$T_{\rm c}$ superconductivity.

To understand the nature of the magnetic correlation, impurity-substitution effects have been studied intensively.
In the case of the substitution of nonmagnetic Zn, muon-spin-relaxation ($\mu$SR) measurements have suggested that dynamical stripe correlations tend to be pinned by Zn and statically stabilized in the underdoped and optimally doped regimes of HTSC.~\cite{Watanabe2002,Adachi2004Zn,Adachi2008,F}
Moreover, the magnetic correlation has been found to be developed by the Zn substitution in the overdoped regime of LSCO,~\cite{Adachi2008,Risdiana2008} Bi$_2$Sr$_2$Ca$_{1-x}$Y$_x$Cu$_2$O$_{8+\delta}$,~\cite{E,J} YBa$_2$Cu$_3$O$_{7-\delta}$ (YBCO),~\cite{D,Watanabe2000} and Bi$_{1.74}$Pb$_{0.38}$Sr$_{1.88}$CuO$_{6+\delta}$.~\cite{Adachi2011}
Elastic neutron-scattering measurements in underdoped and optimally doped LSCO have also revealed that the incommensurate magnetic order is induced by the Zn substitution.~\cite{G,H,I}
In underdoped YBCO, it has been suggested from nuclear-magnetic-resonance measurements that Zn tends to develop the magnetic correlation.~\cite{K,Julien2000}
In the case of the substitution of magnetic Ni, recently, it has been suggested that Ni operates to pin the dynamical stripe correlations to the same degree as Zn,~\cite{Similarity} taking into account the hole-trapping effect of Ni~\cite{Matsuda2006, Machi2003, Tanabe2009, Hiraka2009, Tsutsui2009, Tanabe2010, Suzuki2010} that a hole is localized around a Ni$^{2+}$ ion leading to the formation of the so-called Zhang-Rice doublet~\cite{Hiraka2009}.
This indicates that Ni is no longer regarded as a magnetic impurity but an electrostatic impurity \cite{note} in HTSC.
Therefore, the difference of the substitution effects of nonmagnetic and magnetic impurities has started to attract renewed interest.

Recently, Fujita {\it et al.} have reported from elastic neutron-scattering measurements that incommensurate magnetic and nuclear peaks are observed through the partial substitution of Fe with a large magnetic moment of the spin quantum number, $S$, $= 5/2$ for Cu in LSCO at the hole concentration per Cu, $p, \sim 1/8$,~\cite{I,Fujita2008} suggesting that Fe is effective for the static stabilization of the dynamical stripe correlations more than Zn and Ni.
Moreover, the Fe substitution has been found to induce an incommensurate static magnetic order in the overdoped regime of both LSCO~\cite{RHHe} and Bi$_{1.75}$Pb$_{0.35}$Sr$_{1.90}$CuO$_{4+\delta}$~\cite{Wakimoto2010, Hiraka2010} as well.
Intriguing is that the incommensurate magnetic order observed in the overdoped regime may be different from the stripe order.
That is, a spin density wave (SDW) state due to the Fermi-surface nesting and a spin-glass state due to the Ruderman-Kittel-Kasuya-Yosida (RKKY) interaction have been proposed to be the origins of the incommensurate magnetic order in the 1\% Fe-substituted LSCO~\cite{RHHe} and the 9\% Fe-substituted Bi$_{1.75}$Pb$_{0.35}$Sr$_{1.90}$CuO$_{4+\delta}$,~\cite{Hiraka2010,Wakimoto2010} respectively.
To clarify the magnetic ground state of Fe-substituted HTSC, 
$\mu$SR measurements are suitable, because the frequency range of spin fluctuations sensed by $\mu$SR is lower than that by neutron-scattering.
Therefore, we performed zero-field (ZF) and longitudinal-field (LF) $\mu$SR measurements of partially Fe-substituted La$_{2-x}$Sr$_x$Cu$_{1-y}$Fe$_y$O$_4$ (LSCFO) with $y = 0.005$ and $0.01$ in a wide range of $p$.
In this paper, we report the $\mu$SR results and discuss distinct magnetic states in the underdoped and overdoped regimes of LSCFO.

\section{Experimental}

Polycrystalline samples of LSCFO with $x = 0.06$ -- $0.235$ and $y = 0.005, 0.01$ were prepared by the ordinal solid-state reaction method.~\cite{Adachi2009}
All the samples were checked by the powder x-ray diffraction to be of the single phase.
Electrical-resistivity measurements revealed the good quality of the samples.
ZF- and LF-$\mu$SR measurements were performed at the RIKEN-RAL Muon Facility at the Rutherford-Appleton Laboratory in the UK, using a pulsed positive surface muon beam.
The asymmetry parameter $A(t)$ at a time $t$ was given by
$A(t) = \{ F(t) - \alpha B(t)\}/\{ F(t) + \alpha B(t)\}$,
where $F(t)$ and $B(t)$ are total muon events of the forward and backward counters, which were aligned in the beam line, respectively.
The $\alpha$ is the calibration factor reflecting  the relative counting efficiencies between the forward and backward counters.
The $\mu$SR time spectrum, namely, the time evolution of $A(t)$ was measured at low temperatures down to 2 K.

\section{Results}


\begin{figure*}[hp]
\centering
\includegraphics[height=0.9\textheight]{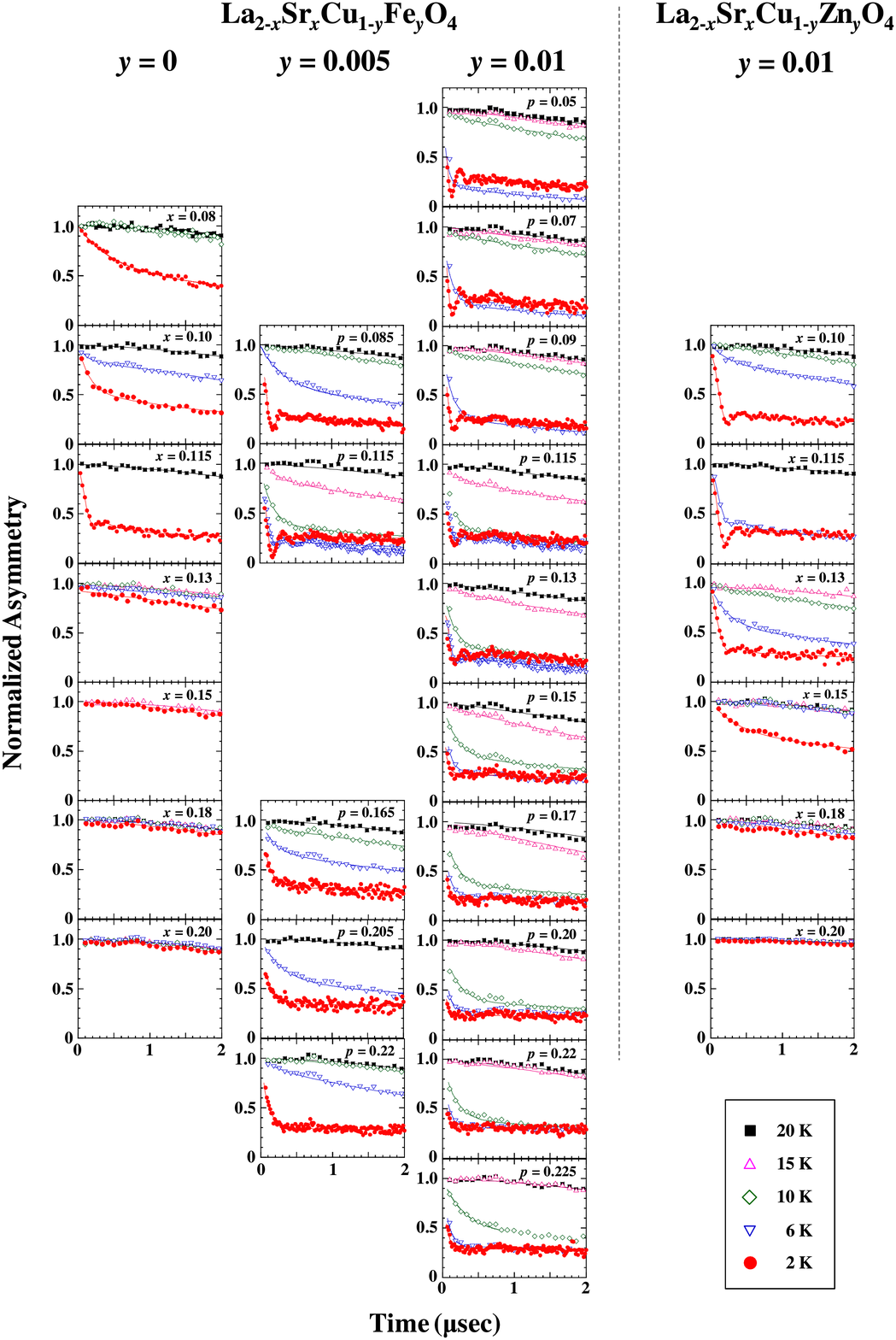}
\caption{ZF-$\mu$SR time spectra in the early time region from 0 to 2 $\mu$sec in La$_{2-x}$Sr$_x$Cu$_{1-y}$Fe$_y$O$_4$ with $y = 0.005$ and $0.01$ at various temperatures down to 2 K.
The hole concentration per Cu, $p$, shown in $y = 0.005$ and $0.01$ is defined as $p = x - y$, considering the Fe$^{3+}$ state.
For comparison, ZF-$\mu$SR time spectra of impurity-free La$_{2-x}$Sr$_x$CuO$_4$ and 1\% Zn-substituted La$_{2-x}$Sr$_x$Cu$_{1-y}$Zn$_y$O$_4$ with $y = 0.01$ are also shown.~\cite{Adachi2008, Risdiana2008, Similarity, Adachi2004Zn}
All the spectra are shown after subtracting the background from the raw spectra and being normalized by the value of the asymmetry at $t = 0$.
Solid lines indicate the best-fit results using $A(t) = A_0 e^{-\lambda _0t}G_Z(\Delta ,t) + A_1e^{-\lambda _1t} + A_2e^{-\lambda _2t}\cos (\omega t + \phi)$.}
\label{NA}
\end{figure*}

\begin{figure*}[tbp]
\centering
\includegraphics[width=\linewidth]{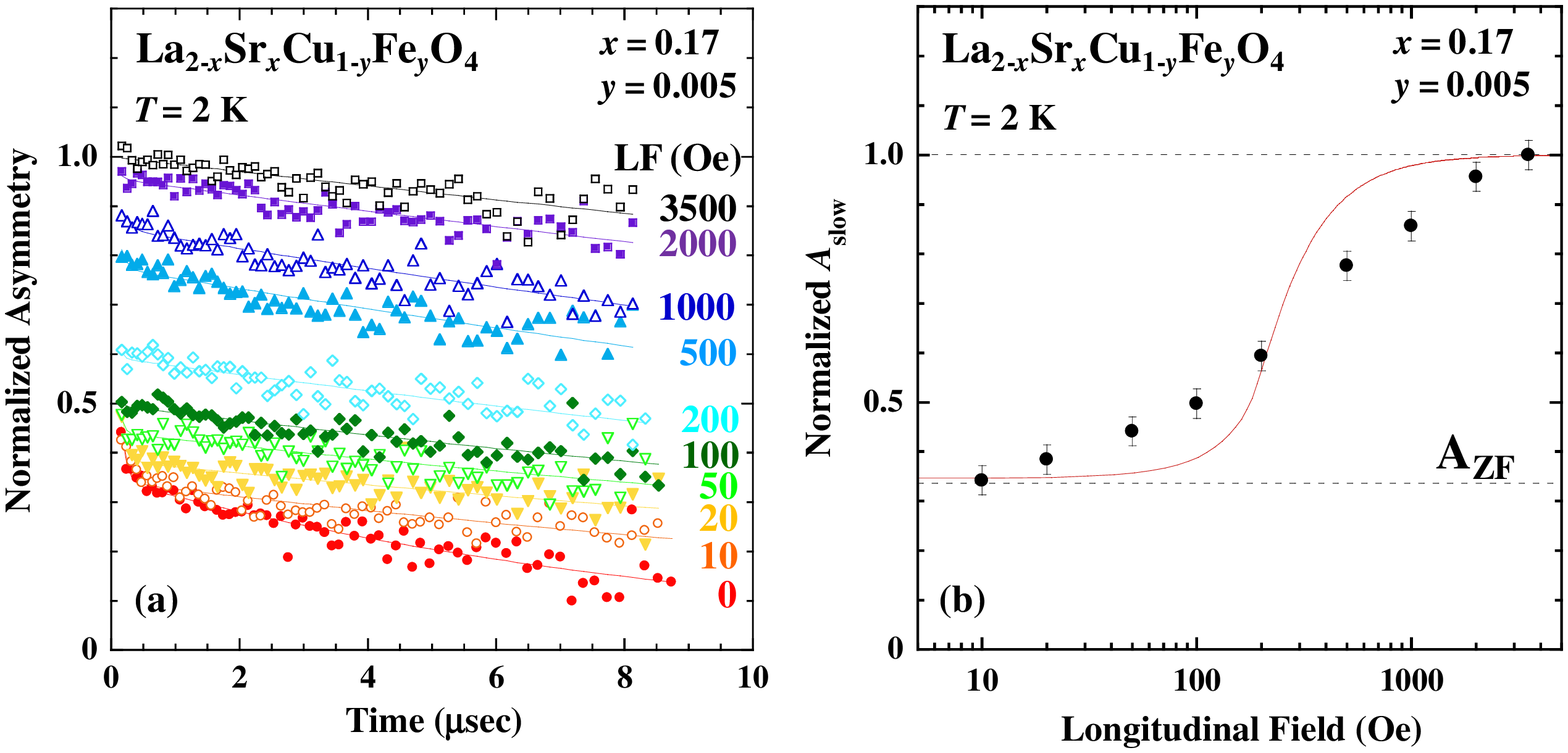}
\caption{(a) Longitudinal-field $\mu$SR time spectra in LF = 0 -- 3500 Oe at 2 K in La$_{2-x}$Sr$_x$Cu$_{1-y}$Fe$_y$O$_4$ with $x = 0.17$ and $y = 0.005$.
Solid lines are the best fit results using
$A^{\rm LF}(t) = A_{\rm slow} e^{-\lambda _{\rm slow} t} + A_{\rm fast} e^{-\lambda _{\rm fast} t}$.
(b) Longitudinal-field dependence of $A_{\rm slow}$ at 2 K in La$_{2-x}$Sr$_x$Cu$_{1-y}$Fe$_y$O$_4$ with $x = 0.17$ and $y = 0.005$.
The solid line represents the best-fit result using
$A_{\rm slow} = \frac{3}{2}(1-A_{\rm ZF})\{ \frac{3}{4} - \frac{1}{4k^2} + \frac{(k^2-1)^2}{8k^3}\log \bigl| \frac{k+1}{k-1}\bigr|\} + \frac{3}{2}(A_{\rm ZF}-\frac{1}{3})$ where $k = \frac{\mu _0H_{\rm LF}}{B_{\rm int}}$.
The lower dashed line represents the zero-field value of $A_{\rm slow}$, $A_{\rm ZF}$.}
\label{LF}
\end{figure*}

Figure \ref{NA} shows the ZF-$\mu$SR time spectra of Fe-substituted LSCFO with $y = 0.005$ and $0.01$ in a wide range of $p$.
For comparison, ZF-$\mu$SR spectra of impurity-free LSCO and 1\% Zn-substituted La$_{2-x}$Sr$_x$Cu$_{1-y}$Zn$_y$O$_4$ (LSCZO) with $y = 0.01$ formerly obtained by our group are also shown.~\cite{Adachi2008,Risdiana2008,Similarity,Adachi2004Zn}
Here, $p$ is defined as $p = x - y$ for LSCFO due to the substitution of trivalent Fe$^{3+}$ for divalent Cu$^{2+}$ and as $p = x$ for LSCO and LSCZO.
All the spectra are shown after subtracting the background from the raw spectra and being normalized by the value of the asymmetry at $t = 0$.
At a high temperature of 20 K, all the spectra show slow depolarization of the Gaussian type due to the nuclear dipole field randomly oriented at the muon site.
This indicates almost no effect of electron spins on the $\mu$SR spectra.
With decreasing temperature, a fast muon-spin depolarization appears in the underdoped regime of LSCO and LSCZO and in the whole $p$ range of LSCFO, indicating a development of the magnetic correlation.
A muon-spin precession is observed in the underdoped regime of LSCFO and LSCZO and at $p \sim 1/8$ of LSCO, though the damping is large.
This indicates the formation of a long-range magnetic order.

It is clearly shown that the muon-spin depolarization is notably enhanced through the Fe substitution in a wide range of $p$.
In the underdoped regime, the muon-spin precession is induced at low temperatures through the Fe substitution due to the formation of a long-range magnetic order, whereas no precession is observed down to 2 K for LSCO with $x < 0.115$.
At $p \sim 0.115$ where a muon-spin precession is observed at 2 K in both of LSCO and LSCFO, both the amplitude and frequency of the precession are enhanced through the Fe substitution, corresponding to the increase of the volume fraction of the magnetically ordered region in the sample and to the increase of the internal magnetic field at the muon site, respectively.
In the overdoped regime of $p \ge 0.15$, though no fast depolarization of muon spins is observed in LSCO nor LSCZO, a fast muon-spin depolarization, followed by an almost flat spectrum with the normalized asymmetry of $\sim 1/3$, is induced at low temperatures through the Fe substitution, indicating the formation of a short-range magnetic order.
This is also evidenced by LF-$\mu$SR at 2 K in overdoped LSCFO with $x = 0.17$ and $y = 0.005$ shown in Fig. \ref{LF}(a).
That is, it is found that the LF spectra shift in parallel up to 3500 Oe, which is a typical decoupling behavior of a static magnetic order.
These results indicate the strong Fe-substitution effect on the development of the magnetic correlation in a wide range of $p$.

In order to obtain the detailed information on the magnetic correlation, the ZF-$\mu$SR time spectra were analyzed using the following standard function:
\begin{equation}
\hspace{-12pt}
A(t) = A_0 e^{-\lambda _0t}G_{\rm Z}(\Delta ,t) + A_1e^{-\lambda _1t} + A_2e^{-\lambda _2t}\cos (\omega t + \phi).
\label{eq1}
\end{equation}
The first term represents the slowly depolarizing component in a region where electron spins fluctuate faster than the $\mu$SR frequency range of $10^6$ -- $10^{11}$ Hz.
The second term represents the fast depolarizing component in a region where the electron-spin fluctuations slow down and/or a short-range magnetic order is formed.
The third term represents the muon-spin precession in a region where a long-range magnetic order is formed.
The $A_0$, $A_1$, $A_2$ and $\lambda_0$, $\lambda_1$ are initial asymmetries and depolarization rates of each component, respectively.
The $G_{\rm Z}(\Delta, t)$ is the static Kubo-Toyabe function describing the muon-spin depolarization due to the nuclear dipole field at the muon site with the distribution width, $\Delta$.~\cite{Uemura1985}
The $\lambda_2$, $\omega$, $\phi$ are the damping rate, frequency and phase of the muon-spin precession, respectively.
The time spectra are well fitted with Eq. (\ref{eq1}), as shown by solid lines in Fig. \ref{NA}.


\begin{figure}[tbp]
\includegraphics[width=0.95\linewidth]{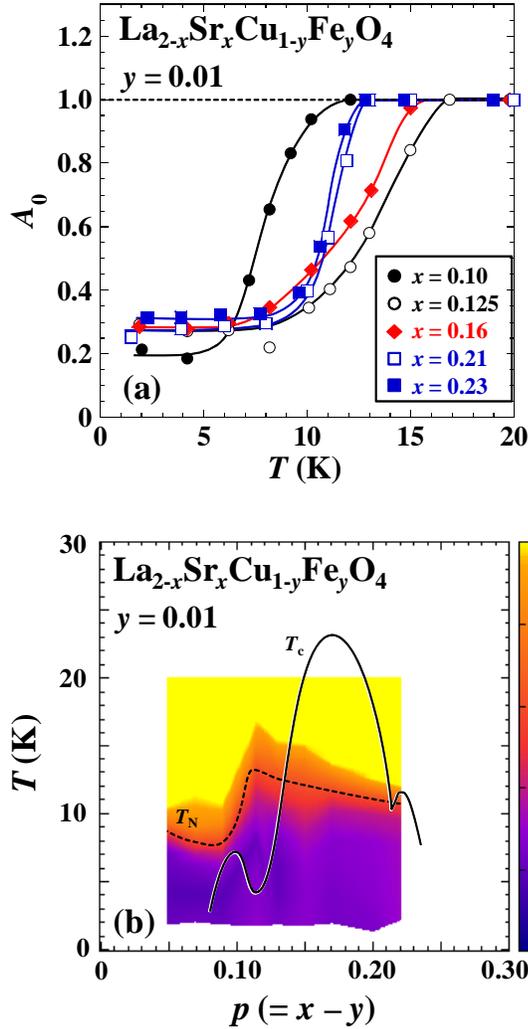}
\caption{(a) Tempeature dependence of the initial asymmetry of the slowly depolarizing component, $A_0$, in Eq. (\ref{eq1}) normalized by its value at 20 K for typical values of $x$ in La$_{2-x}$Sr$_x$Cu$_{1-y}$Fe$_y$O$_4$ with $x = 0.01$.
Solid lines are to guide the reader's eye.
(b) Contour map of the change in $A_0$ with temperature and hole concentration, $p = x -y$, in La$_{2-x}$Sr$_x$Cu$_{1-y}$Fe$_y$O$_4$ with $x = 0.01$.
Solid and dashed lines represent the superconducting transition temperature, $T_{\rm c}$, defined at the midpoint of the superconducting transition in the electrical resistivity and the magnetic transition temperature, $T_{\rm N}$, estimated from the temperature dependence of $A_0$ in Eq. (\ref{eq1}) for La$_{2-x}$Sr$_x$Cu$_{1-y}$Fe$_y$O$_4$ with $y = 0.01$, respectively.}
\label{fig3}
\end{figure}

Figures \ref{fig3}(a) and (b) show the temperature dependence of $A_0$ normalized by its value at a high temperature of 20 K and a contour map of the change in $A_0$ with temperature and $p$ in LSCFO with $y = 0.01$, respectively.
The change in $A_0$ is often used as a probe of the magnetic transition, because it reflects the volume fraction of the nonmagnetic region.~\cite{Watanabe1992}
At $\sim$ 20 K, $A_0 = 1$ in an entire range of $p$, indicating that electron spins fluctuate fast beyond the $\mu$SR frequency range.
With decreasing temperature, $A_0$ tends to decrease due to the appearance of the fast depolarizing component corresponding to the development of the magnetic correlation, and reaches $\sim 1/3$ at low temperatures, indicating the formation of a static magnetic order.
It is found that $A_{\rm 0}$ starts to decrease at the highest temperature at $p \sim 1/8$ among all the samples, indicating that the magnetic transition temperature exhibits the maximum at $p \sim 1/8$.
This is consistent with the behavior of LSCO and LSCZO where the stripe order is most stable at $p \sim 1/8$.~\cite{Adachi2004Zn}
As for the temperature width of the magnetic transition, it is found that a zonal intermediate region of $A_{\rm 0}$ ranging from 1 to $\sim$ 1/3 is rather wide in temperature in the underdoped regime below $p \sim 0.15$, while it is narrow in the overdoped regime above $p \sim 0.15$.
These suggest that the magnetic transitions are broad and sharp below and above $p \sim 0.15$, respectively.
It is noted that the broad magnetic transition is typical of the impurity-induced formation of the stripe order.~\cite{Adachi2004Zn,Adachi2008,G,H,Kimura1999}
Therefore, this result suggests a possible change of the magnetic nature between the underdoped and overdoped regimes.


\begin{figure}
\centering
\includegraphics[width=\linewidth]{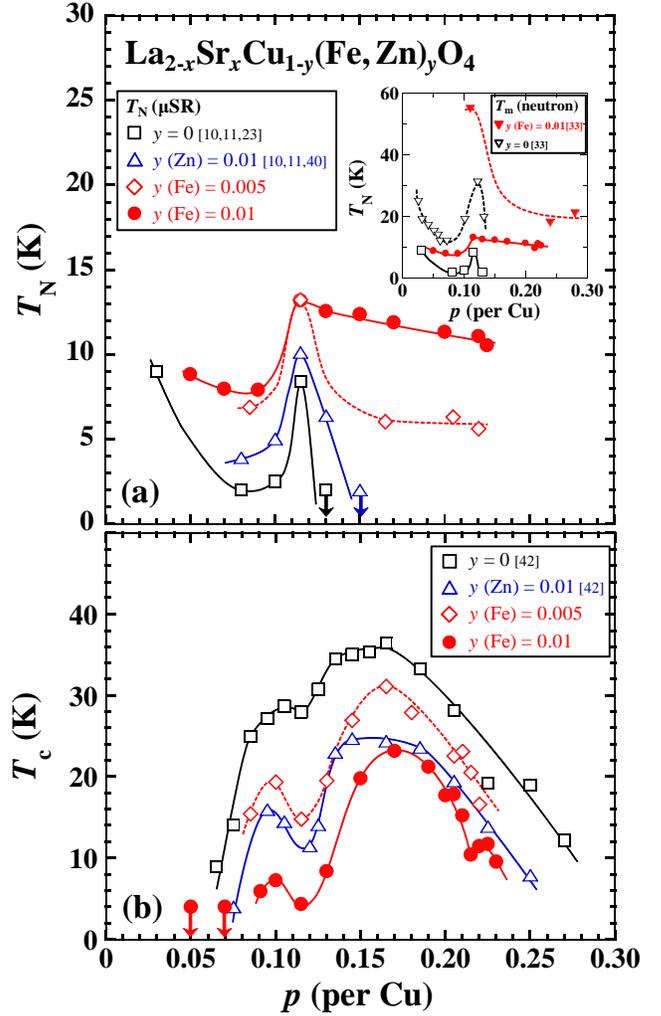}
\caption{(a) Hole-concentration, $p$, dependence of the magnetic transition temperature, $T_{\rm N}$, estimated from the temperature dependence of $A_0$ in Eq. (\ref{eq1}) for La$_{2-x}$Sr$_x$Cu$_{1-y}$Fe$_y$O$_4$ (LSCFO) with $y = 0.005$ and $0.01$.
The data of impurity-free La$_{2-x}$Sr$_x$CuO$_4$ (LSCO) and 1\% Zn-substituted La$_{2-x}$Sr$_x$Cu$_{1-y}$Zn$_y$O$_4$ (LSCZO) with $y = 0.01$ are also plotted for comparison.~\cite{Adachi2008, Similarity, Adachi2004Zn, Suzukiunpublish}
Inset shows the magnetic transition temperature estimated from neutron-scattering measurements, $T_{\rm m}$.~\cite{RHHe}
(b) $p$ dependence of the superconducting transition temperature, $T_{\rm c}$, defined at the midpoint of the superconducting transition in the electrical resistivity in LSCFO with $y = 0.005, 0.01$.
The data of impurity-free LSCO and 1\% Zn-substituted LSCZO with $y = 0.01$ are also plotted for comparison.~\cite{Koike1992}
Values of $p$ are defined as $p = x - y$ for LSCFO and as $p = x$ for LSCO and LSCZO.
Solid and dashed lines are to guide the reader's eye.}
\label{TNTc}
\end{figure}

Figure \ref{TNTc}(a) shows the $p$ dependence of the magnetic transition temperature, $T_{\rm N}$, defined at the midpoint of the change in $A_0$ from unity to the averaged value in the static magnetic state at low temperatures in each composition.
Values of $T_{\rm N}$ for impurity-free LSCO and 1\% Zn-substituted LSCZO estimated in the similar way are also plotted for comparison.~\cite{Adachi2008, Similarity, Adachi2004Zn, Suzukiunpublish}
In the underdoped regime of $p \lesssim 1/8$, it is apparent that the Fe substitution raises $T_{\rm N}$ more markedly than the Zn substitution.
However, the $p$ dependence of $T_{\rm N}$ exhibits a qualitatively similar behavior regardless of the kind of impurities.
That is, $T_{\rm N}$ increases toward $p = 0$ where the three-dimensional antiferromagnetic order is formed and exhibits the local maximum at $p \sim 1/8$.
On the other hand, in the overdoped regime of $p \gtrsim 0.15$ where the magnetic correlation weakens and $T_{\rm N}$ no longer exists in LSCO and LSCZO, $T_{\rm N}$ appears through the Fe substitution.
Surprisingly, the decrease in $T_{\rm N}$ with increasing $p$ is gradual for LSCFO with $y = 0.01$, resulting in higher values of $T_{\rm N}$ than those in the underdoped regime.
These results disagree with the general understanding that the magnetic correlation tends to weaken markedly with increasing $p$ in the overdoped regime.~\cite{Kastner1998}
The robust $T_{\rm N}$ against hole doping for $p > 1/8$ is in contrast to the significant decrease in the onset temperature of the magnetic transition estimated from elastic neutron-scattering measurements, $T_{\rm m}$, with hole doping, as shown in the inset of Fig. \ref{TNTc}(a).~\cite{RHHe}
As a result, values of $T_{\rm N}$ and $T_{\rm m}$ are close to each together in the overdoped regime.
Given the difference of the frequency range between $\mu$SR and neutron scattering, the close values between $T_{\rm N}$ and $T_{\rm m}$ indicate a sharp magnetic transition, which is consistent with the narrow temperature-width of the magnetic transition seen in Fig. \ref{fig3}.
In addition, it is found that $T_{\rm N}$ exhibits a significant dependence on the Fe concentration in the overdoped regime, whereas $T_{\rm N}$ is saturated above $y \ge 0.005$ for $p < 1/8$.

As for the Fe-substitution effect on the superconductivity, the $p$ dependence of $T_{\rm c}$, defined at the midpoint of the SC transition in the electrical resistivity, is shown in Fig. \ref{TNTc}(b) for LSCFO as well as for impurity-free LSCO and 1\% Zn-substituted LSCZO.~\cite{Koike1992}
It is found in the underdoped regime that a dip of $T_{\rm c}$ is observed at $p \sim 1/8$ both in LSCFO with $y = 0.005$ and $0.01$  and that $T_{\rm c}$ is strongly depressed through the Fe substitution,~\cite{L} concomitant with the enhancement of $T_{\rm N}$ as shown in Fig. \ref{TNTc}(a), which is characteristic of the competitive relation between the stripe order and superconductivity.
In the overdoped regime, on the other hand, $T_{\rm c}$ is found to be maintained at high temperatures, while $T_{\rm N}$ is also rather high.
These contrasting results suggest that the nature of the magnetic order induced by the Fe substitution in the overdoped regime might be different from that in the underdoped regime.


\begin{figure}[tbp]
\includegraphics[width=\linewidth]{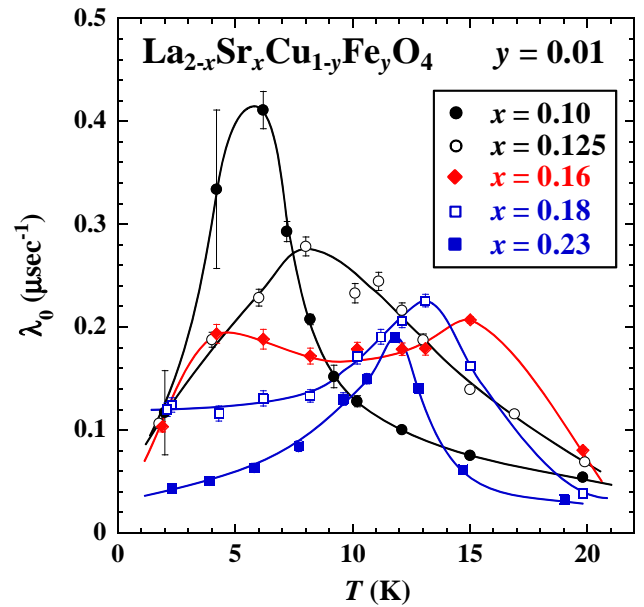}
\caption{Temperature dependence of the depolarization rate of the slowly depolarizing component, $\lambda _0$, for typical values of $x$ in La$_{2-x}$Sr$_x$Cu$_{1-y}$Fe$_y$O$_4$ with $y = 0.01$.
Solid lines are to guide the reader's eye.}
\label{lambda0}
\end{figure}

Another characteristic of the magnetic transition in LSCFO with $y = 0.01$ is able to be seen in the temperature dependence of $\lambda _0$ shown in Fig. \ref{lambda0}.
In general, $\lambda _0$ increases with decreasing temperature toward $T_{\rm N}$ due to the critical slowing down, followed by the decrease below $T_{\rm N}$ due to the frequency of the magnetic fluctuation being lower than the $\mu$SR frequency range.
An apparent peak is observed at $\sim$ 6 K and $\sim$ 8 K in underdoped samples of $x = 0.10$ and $0.125$, respectively, while in overdoped samples of $x = 0.18$ and $0.23$ a peak is observed at higher temperature of $\sim$ 12.5 K than that in the underdoped regime.
Intriguingly, $\lambda_0$ exhibits two peaks in the intermediate sample of $x = 0.16$, suggesting that there exist two kinds of magnetic transition in this sample; one has the nature of the underdoped magnetism and the other has that of the overdoped one.


\begin{figure}[tbp]
\centering
\includegraphics[width=0.95\linewidth]{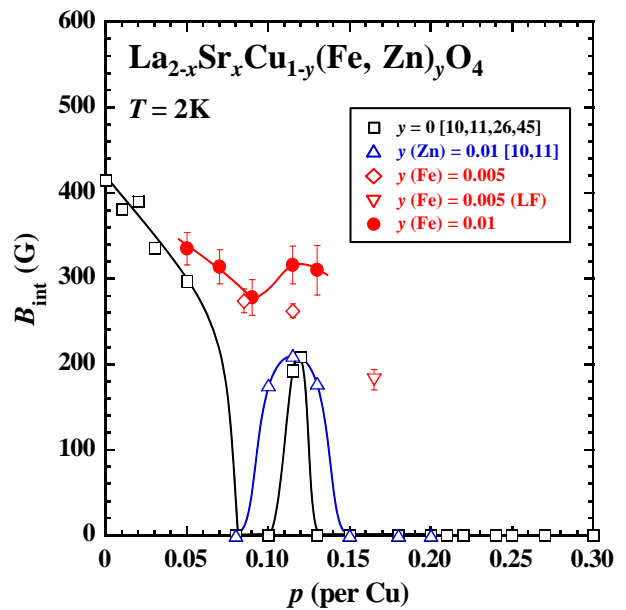}
\caption{Hole-concentration, $p$, dependence of the internal magnetic field at the muon site, $B_{\rm int}$, estimated from the frequency of the muon-spin precession in the ZF-$\mu$SR spectra at 2 K in La$_{2-x}$Sr$_x$Cu$_{1-y}$Fe$_y$O$_4$ with $y = 0.005$ and $0.01$.
The open inverted-triangle is $B_{\rm int}$ estimated from the LF-$\mu$SR spectra at 2 K.
For comparison, the data of impurity-free La$_{2-x}$Sr$_x$CuO$_4$~\cite{Adachi2004Zn, Adachi2008, Tanabe2009, Budnick1988} and 1\% Zn-substituted La$_{2-x}$Sr$_x$Cu$_{1-y}$Zn$_y$O$_4$ with $y = 0.01$~\cite{Adachi2008, Adachi2004Zn} are also plotted.
Solid lines are to guide the reader's eye.}
\label{Bint}
\end{figure}

Here, the internal magnetic field at the muon site, $B_{\rm int}$, in LSCFO is discussed.
The $B_{\rm int}$ is generally obtained directly from $\omega$ as $B_{\rm int} = \mu _0\omega/\gamma _\mu$ where $\gamma _{\mu}$ is the gyromagnetic ratio of muon spin ($\gamma _{\mu}/2\pi = 13.55$ MHz/kOe) and $\mu _0$ is the space permeability.
Figure \ref{Bint} shows the $p$ dependence of $B_{\rm int}$.
The data of impurity-free LSCO and 1\% Zn-substituted LSCZO in literatures are also plotted.~\cite{Adachi2008, Adachi2004Zn, Tanabe2009, Budnick1988}
In LSCO, $B_{\rm int}$ decreases with increasing $p$, shows the local maximum at $p \sim 1/8$ and disappears above $p > 0.13$.
It is found that $B_{\rm int}$ is enhanced in a wide range of $p$ through the Fe substitution and higher than $B_{\rm int}$ of 1\% Zn-substituted LSCZO, indicating that the magnetic order is more stabilized through the Fe substitution than through the Zn substitution.

As for the overdoped regime, $B_{\rm int}$ is unable to be estimated from the ZF-spectra, because no muon-spin precession was observed.
Instead, $B_{\rm int}$ can be estimated from the decoupling spectra in LF spectra in Fig. \ref{LF}(a), using the following two exponential function:
\begin{equation}
A^{\rm LF}(t) = A_{\rm slow}e^{-\lambda_{\rm slow} t} + A_{\rm fast}e^{-\lambda _{\rm fast} t}.
\end{equation}
The first and second terms represents the slowly and fast depolarizing components above and below $\sim 1$ $\mu$sec in the spectrum, respectively.
The $A_{\rm slow}$ and $A_{\rm fast}$ are initial asymmetries and $\lambda _{\rm slow}$ and $\lambda _{\rm fast}$ are the depolarization rates of each component.
The LF dependence of $A_{\rm slow}$ at 2 K is shown in Fig. \ref{LF}(b).
The lower dashed line represents the zero-field value of $A_{\rm slow}$, $A_{\rm ZF}$.
It is found that $A_{\rm slow}$ increases gradually with increasing LF, corresponding to the decoupling behavior of the polarization of muon spins.
The solid line in Fig. \ref{LF}(b) indicates the fitting result using the following decoupling function:~\cite{Pratt2007}
\begin{multline}
\label{decoupling_function}
A_{\rm slow} = \frac{3}{2}(1-A_{\rm ZF})\{ \frac{3}{4} - \frac{1}{4k^2} + \frac{(k^2-1)^2}{8k^3}\log \biggl| \frac{k+1}{k-1}\biggr|\} \\
+ \frac{3}{2}(A_{\rm ZF}-\frac{1}{3}),
\end{multline}
where $k$ is the ratio of LF to the internal field, $\mu _0H_{\rm LF}/B_{\rm int}$.
Note that this function assumes that muons feel a single value of $B_{\rm int}$ which is randomly oriented in the polycrystalline sample.
The $B_{\rm int}$ of LSCFO with $x = 0.17$ and $y = 0.005$ is estimated to be 182 G.
However, the recovery of $A_{\rm slow}$ to the unity is broader against LF than that expressed by Eq.(\ref{decoupling_function}), suggesting that $B_{\rm int}$ at the muon site is not uniform.
This is reasonable, for the magnetic order is rather short-ranged.

\section{Discussion}\label{Discussion}


The present results clearly indicate that the magnetic order is formed through only 1\% Fe substitution for LSCFO in a wide range of $p$ where superconductivity appears in LSCO.
The Fe substitution enhances $T_{\rm N}$ and $B_{\rm int}$  more markedly than the Zn substitution in the underdoped regime.
Moreover, the static magnetic order appears through the Fe substitution even in the overdoped regime where no development of the magnetic correlation is observable in LSCO from $\mu$SR measurements.~\cite{Adachi2008,Risdiana2008}
On the other hand, the suppression of superconductivity by the Fe substitution is larger than by the Zn substitution especially in the underdoped regime.
Given the large magnetic moment of Fe, this evokes conventional superconductors based on the BCS theory~\cite{Tinkham} where $T_{\rm c}$ is depressed by magnetic impurities more than by nonmagnetic impurities.
However, the present results of LSCFO suggest that a novel magnetic state is in deep relation to the superconductivity, being discussed in the followings.


In the underdoped regime, both $T_{\rm N}$ and $B_{\rm int}$ show the local maximum at $p \sim 1/8$ in correspondence to the local minimum of $T_{\rm c}$.
Moreover, the broad magnetic transition suggests that the magnetically ordered region is created around impurities and expands throughout the sample with decreasing temperature.
These are characteristics of the formation of the stripe order stabilized around impurities.
Therefore, the magnetic order in underdoped LSCFO is understood to be due to the pinning of the dynamical stripe correlations by Fe.

It is speculated that the strong pinning effect of Fe is due to the large magnetic moment of Fe$^{3+}$, namely, $S = 5/2$ suppressing the stripe fluctuations.
One possible reason for the significant effect of magnetic impurities on the static stabilization of the stripe correlations is that the difference of the free energy between the case of Fe being located in the spin domain of the stripes and the case of Fe being located in the charge domain of the stripes is large, taking into consideration both the exchange interaction between spins and the transfer integral of holes.
The other possible reason is that the large magnetic moment of Fe$^{3+}$ tends to induce a static magnetic order as in the case of the application of external magnetic field.
In fact, the dynamical stripe correlations are statically stabilized by the application of magnetic field parallel to the c-axis.~\cite{Lake2002,Kudo2004, Adachi2005, Chang2008}


On the other hand, the magnetic order induced by the Fe substitution in the overdoped regime involves following characteristics different from those in the underdoped regime: (i) the absence of muon-spin precession as shown in Fig. \ref{NA}, (ii) the sharp magnetic transition as shown in Fig. \ref{fig3}, (iii) the large $T_{\rm N}$ values in spite of the magnetic correlation weakened with hole doping, (iv) the marked enhancement of $T_{\rm N}$ by the Fe substitution, and (v) the robust high $T_{\rm c}$ values in spite of the large $T_{\rm N}$ values.
Supposed the stripe-pinning effect by Zn in overdoped LSCZO,~\cite{Risdiana2008} the stripes are probably disordered due to excess holes overflowing into the spin domain and therefore causing a disorder in the magnetic state.
This is consistent with the absence of muon-spin precession.
This concept of the stripe pinning, however, appears to be incompatible with higher $T_{\rm N}$ values in the overdoped regime  than in the underdoped regime, because the stripe correlations weaken progressively with overdoping.~\cite{Wakimoto2004}

A simple SDW state driven by the nesting of the Fermi surface, suggested by He {\it et al.},~\cite{RHHe} is a possible candidate to explain the peculiar magnetic state in overdoped LSCFO.
Combined experiments of elastic neutron-scattering and angle-resolved photoemission spectroscopy in overdoped LSCFO have revealed that the 1\% Fe substitution induces incommensurate magnetic peaks around ($\pi$/2, $\pi$/2) in the reciprocal lattice space and that the incommensurability is larger than 1/8.
As the incommensurability is well reproduced in terms of the Fermi-surface nesting, it has been suggested that the incommensurate magnetic order is driven by the Fermi-surface instability in the overdoped regime, while the stripe order is stabilized by Fe in the underdoped regime.
This scenario is consistent with the present results in the viewpoint that the origin of the magnetic state is different between the underdoped and overdoped regimes.
This is, however, unable to explain the absence of muon-spin precession in the $\mu$SR spectrum nor the strong dependence of $T_{\rm N}$ on the Fe concentration in overdoped LSCFO.

The magnetic order due to the RKKY interaction between Fe spins is also a candidate for the magnetic state in overdoped LSCFO.
A typical example of the RKKY interaction is seen in dilute magnetic alloys such as Cu-Mn, in which dilute magnetic moments form a spin-glass state mediated by itinerant electrons.~\cite{Uemura1985,Lamelas1995}
In cuprates, early studies from magnetic-susceptibility and electron-spin-resonance measurements have suggested that a spin-glass state due to the RKKY interaction is induced by the Fe substitution in optimally doped LSCO.~\cite{Cieplak1993}
Moreover, Hiraka {\it et al.} have recently proposed that a static incommesurate magnetic order induced by the 9\% Fe substitution in overdoped Bi$_{1.75}$Pb$_{0.35}$Sr$_{1.90}$CuO$_{4+\delta}$ is related to the RKKY interaction,~\cite{Hiraka2010} which is supported by magnetic-field effects of the neutron scattering and electrical resistivity.~\cite{Wakimoto2010}
In this case, the fast depolarization without precession in the $\mu$SR spectra in overdoped LSCFO is well explained as being due to the randomly distributed magnetic moments.~\cite{Uemura1985}
Moreover, the present result that $T_{\rm N}$ increases linearly with increasing $y$ in overdoped LSCFO as shown in Fig. \ref{TNTc}(a) is well explained also.
These results indicate that a Fermi-liquid-like ground state is realized in overdoped LSCO, as revealed from the specific-heat and electrical-resistivity measurements in the overdoped regime of Ni-substituted La$_{2-x}$Sr$_x$Cu$_{1-y}$Ni$_y$O$_4$.~\cite{Tanabe2010,Suzuki2010}
That is, the electronic state based upon the strong electron correlation in the underdoped regime might change to be Fermi-liquid-like in the overdoped regime.

The change of the two magnetic states with changing hole-doping is able to be guessed from Fig. \ref{lambda0}.
It appears that two peaks of $\lambda _0$ observed in $x = 0.16$ originate from the stripe order of Cu spins at low temperatures below $\sim$ 4 K and from the spin-glass state of Fe spins at temperatures below $\sim$ 15 K.
Therefore, it is guessed that the magnetic state shows a crossover-like change with hole doping through a coexisting regime consisting of the stripe order characteristic of the underdoped regime and the spin-glass state of Fe spins  characteristic of the overdoped regime.


Finally, we briefly discuss Fe-substitution effects on the nano-scale phase separation into SC and normal metallic regions in the overdoped regime of LSCO, which has been suggested from $\mu$SR,~\cite{Uemura2003} magnetic-susceptibility~\cite{Tanabe2005,Tanabe2007} and specifit-heat measurements.~\cite{Loram1994,Wang2007}
In the phase-seprated overdoped regime, it seems that the magnetic correlation due to Cu spins still remains in the SC region and that the normal-state region is rather metallic.
Assuming the phase separation persists in overdoped LSCFO, it is speculated that the RKKY interation occurs in the metallic region.
In fact, however, the normalized asymmetry of ZF-$\mu$SR spectra in the overdoped regime decreases to the value of $\sim 1/3$ as shown in Fig. \ref{NA}, indicating all muon spins stopping in the sample depolarize fast.
This suggests that a spin-glass state is realized at low temperatures in overdoped LSCFO irrespective of the phase separation.

\section{Summary}

In order to investigate Fe-substitution effects on the magnetic correlation in LSCO, we have carried out ZF- and LF-$\mu$SR measurements in a wide range of $p$ of LSCFO with $y = 0.005$ and $0.01$.
The ZF-$\mu$SR spectra have clearly revealed that the magnetic correlation is developed through the 1\% Fe substitution in a wide range of $p$ more markedly than through the 1\% Zn substitution.
Both the broad magnetic transition in the underdoped regime and the local maximum of $T_{\rm N}$ in correspondence to the local minimum of $T_{\rm c}$ at $p \sim 1/8$ suggest that the magnetic order induced by the Fe substitution is due to the stripe pinning by Fe and that magnetic impurities are more effective for the static stabilization of the dynamical stripe correlations than nonmagnetic impurities.
On the other hand, the Fe-induced static magnetic order in the overdoped regime has included the following characteristics different from those in the underdoped regime: (i) the absence of muon-spin precession, (ii) the sharp magnetic transition, (iii) the large $T_{\rm N}$ values in spite of the magnetic correlation weakened with hole doping, (iv) the marked enhancement of $T_{\rm N}$ by the Fe substitution, and (v) the robust high $T_{\rm c}$ values in spite of the large $T_{\rm N}$ values.
It has been concluded that the spin-glass-like magnetic state due to the RKKY interaction is most likely and that the pristine ground state in LSCO is Fermi-liquid-like in the overdoped regime.
That is, the electronic ground state in LSCO has been concluded to change crossover-like from the localized state with the stripe correlations in the underdoped regime to the itinerant Fermi-liquid state in the overdoped regime.

\section*{Acknowledgments}
We would like to thank M. Fujita and R.-H. He for their helpful discussion.
The $\mu$SR measurements at RAL were partially supported by the KEK-MSL Inter-University Program for Oversea Muon Facilities and also by the Global COE Program “Materials Integration (International Center of Education and Research), Tohoku University,” of the Ministry of Education, culture, Sports, Science and Technology, Japan.



\begin{thebibliography}{00}
\bibitem{note}
A Ni$^{2+}$ ion with a ligand hole is roughly regarded as a Ni$^{3+}$ ion.
\end{thebibliography}
\end{document}